\documentclass[pra,showkeys,aps,preprint]{revtex4-1}
\usepackage{graphicx,color}

\begin{document}

\title{Proximity induced spin-valley polarization in silicene/germanene on F-doped WS$_2$}
\author{Shahid Sattar, Nirpendra Singh,}
\author{Udo Schwingenschl\"{o}gl}
\email{udo.schwingenschlogl@kaust.edu.sa, +966(0)544700080}
\affiliation{King Abdullah University of Science and Technology (KAUST), Physical
Science and Engineering (PSE) Division, Thuwal 23955-6900, Saudi Arabia}


\begin{abstract}
Silicene and germanene are key materials for the field of valleytronics.
However, interaction with the substrate, which is
necessary to support the electronically active medium, becomes a major obstacle. In the
present work, we propose a substrate (F-doped WS$_2$) that avoids detrimental effects
and at the same time induces the required valley polarization, so that no further steps
are needed for this purpose. The behavior is explained by proximity effects on
silicene/germanene, as demonstrated by first-principles calculations. Broken inversion
symmetry due to the presence of WS$_2$ opens a substantial band gap in silicene/germanene.
F doping of WS$_2$ results in spin polarization, which, in conjunction with
proximity-enhanced spin orbit coupling, creates sizable spin-valley polarization.
\end{abstract}

\maketitle

\section{Introduction}
Silicene and germanene are topological insulators with nontrivial band gaps of 2 meV and
24 meV, respectively, induced by spin orbit coupling (SOC) \cite{Fv}. The band gap can
be controlled electrically by applying a gate voltage in the out-of-plane direction
\cite{tune}. It is anticipated that both materials host quantum spin Hall \cite{qshe},
quantum anomalous Hall \cite{qahe}, and valley polarized quantum anomalous Hall
\cite{vpqahe} phases. Silicene has been prepared on various substrates \cite{ex1,ex2,si_zrb2,a1},
while it is questionable whether it can exist in freestanding form (which also applies
to germanene). This is the reason why various theoretical results on the interaction
with possible substrates are found in the literature, including the insulator $h$-BN
\cite{hbn1}, the semiconductor GaAs \cite{sigaas}, and the metals Ca \cite{simetal}, Ag
\cite{ag}, and Ir \cite{ir}. While on metallic substrates the Dirac behavior of silicene
typically is not maintained \cite{subst1,subst2}, transition metal dichalcogenides are
characterized by a weak interaction \cite{simos2}. From a different
perspective, the latter class of materials is receiving great interest in recent days
due to the fact that it realizes band gaps in a technologically attractive range
\cite{tmdc}. Transition metal dichalcogenides also have proven to be suitable hosts for
graphene \cite{grap1,grap2} and it has been demonstrated that the SOC of graphene can
be enhanced by three orders of magnitude to about 17 meV on WS$_2$ due to proximity
effects \cite{soprox}.

Valleytronics is emerging as a new and exciting area of research, aiming to
exploit the valley degree of freedom in Dirac materials \cite{valley1,valley2}. An
essential prerequisite of valleytronics, of course, is the availability of materials
with valley polarization, i.e., the energetical degeneracy of the valleys at the high
symmetry K and K$'$ points of the hexagonal Brillouin zone must be lifted. While this
is difficult to realize in graphene, the stronger SOC and buckled lattice of
silicene/germanene provide an avenue to access and control the valley degree of
freedom \cite{tronics}. Spin and valley polarization can be achieved by means of doping
and decoration with certain $3d$ or $4d$ transition metals \cite{tm3d,tm4d} as well as
by an external electric field \cite{ef,qahe,vpqahe}. However, interaction with the
substrate is typically detrimental, because the electronic states are perturbed
\cite{subst3}. Besides the need to reduce the interaction with the substrate, it would
be a great advantage if the substrate itself can be used to induce the required valley
polarization, in order to reduce the complexity of the system. In this context, we show
in the present work that proximity effects between silicene/germanene and WS$_2$ can be
utilized to obtain a suitable platform to explore spin and valley physics. We first
discuss the band characteristics and spin splitting in silicene/germanene induced by
the strong SOC in WS$_2$ and afterwards demonstrate that the spin polarization in
F-doped WS$_2$ generates spin-valley polarization.

\begin{figure}[b]
\centering
\includegraphics[width=0.5\textwidth,clip=true]{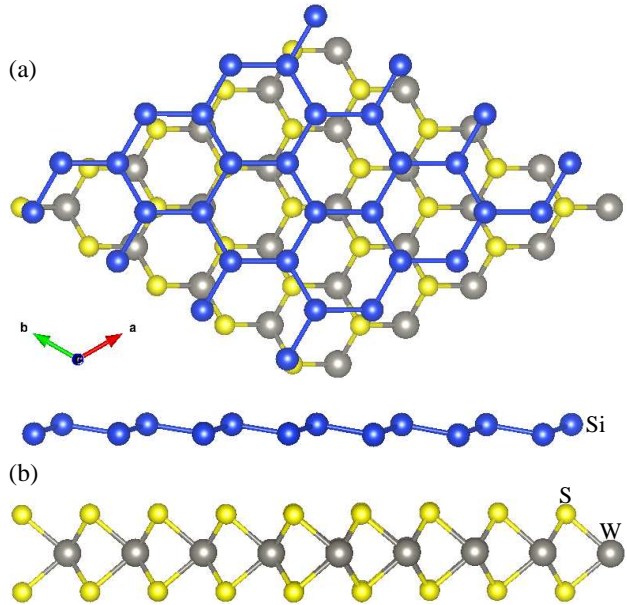}
\caption{Silicene on monolayer WS$_2$: (a) top view and (b) side view.}
\label{fig:fig1}
\end{figure}

\section{Computational method}
We use the Vienna Ab-initio Simulation Package to perform first-principles calculations
based on density functional theory \cite{vasp}. The exchange correlation potential is
described in the generalized gradient approximation, using the Perdew-Burke-Ernzerhof
scheme, and the plane wave cutoff energy is set to a sufficiently large value of 475 eV.
Moreover, the SOC is taken into account in all calculations and the van der Waals
interaction is incorporated using the DFT-D3 method \cite{dftd3}. The
optimized lattice constants of silicene, germanene, and WS$_2$ are 3.86 \AA, 4.05 \AA,
and 3.18 \AA. In order to reduce the
lattice mismatch to 2.9\% and 1.9\%, respectively, a $4\times4\times1$ supercells of
silicene and germanene are placed on top of a $5\times5\times1$ supercell of WS$_2$.
Vacuum slabs of 15 \AA\ thickness are used to obtain two-dimensional models. For the
Brillouin zone integration, Monkhorst-Pack $1\times1\times1$ and $3\times3\times1$
k-meshes are employed in the structure relaxations and band structure calculations,
respectively. We achieve in each case at least an energy convergence of $10^{-6}$
eV and an atomic force convergence of $10^{-2}$ eV/\AA.

\section{Results and Discussion}
The optimized crystal structure of silicene on top of monolayer WS$_2$ is illustrated
in Fig.\ \ref{fig:fig1}. A corresponding figure for germanene looks very similar and
therefore is not shown. For different stackings of silicene/germanene on top of WS$_2$
(different lateral shifts) we obtain very small (few meV) total energy differences,
showing that the materials can easily slide on each other. The distance between the
two component materials turns out to be 3.13 \AA\ in the case of silicene and 2.90 \AA\
in the case of germanene. While the buckling of silicene is hardly affected by the
interaction with WS$_2$ (0.46 \AA), it is slightly enhanced to 0.74 \AA\ for germanene
(0.65 \AA\ in the freestanding case). The smaller interlayer distance and the
enhancement of the buckling demonstrate that WS$_2$ interacts more with germanene than
with silicene, though the coupling is still weak. In order to quantify the interaction,
we calculate the binding energy
$(E_{\rm silicene/germanene+WS_2}-E_{\rm silicene/germanene}-E_{\rm WS_2})/N$,
given by the total energies of the combined system, freestanding silicene/germanene,
and freestanding monolayer WS$_2$. Furthermore, $N$ is the number of Si/Ge atoms. We
obtain values of $-158$ meV and $-171$ meV for silicene and germanene, respectively,
confirming our conclusion that the interaction is stronger in the latter case.

\begin{figure}[t]
\centering
\includegraphics[width=0.5\textwidth,clip=true]{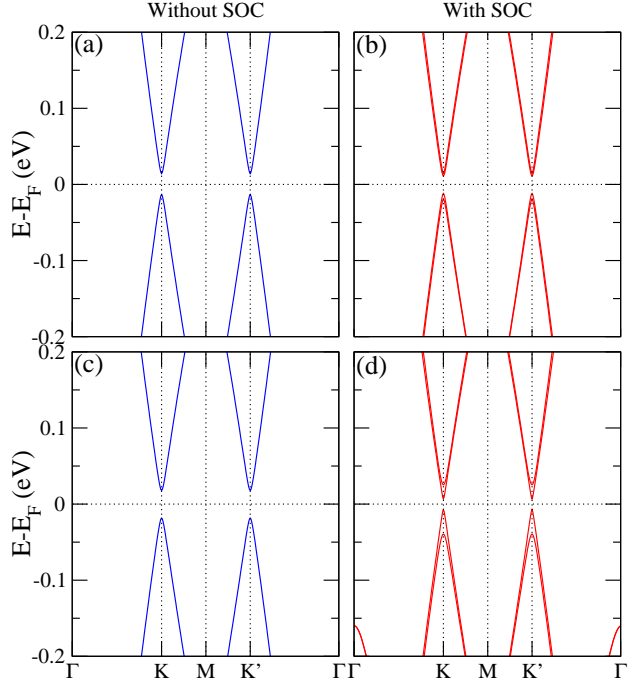}
\caption{Band structures of (a/b) silicene and (c/d) germanene on monolayer WS$_2$.
The SOC is neglected (left) or taken into account (right).}
\label{fig:fig2}
\end{figure}

The electronic band structures in Fig.\ \ref{fig:fig2} show that both silicene and
germanene on WS$_2$ maintain a linear dispersion of the $\pi$ bands in the vicinity
of the Fermi energy. This is the case both when the SOC is neglected and when it is
taken into account. Without SOC we obtain band gaps of 29 meV and 38 meV, see
Fig.\ \ref{fig:fig2}(a/c), for silicene and germanene on WS$_2$, respectively, which is
comparable to the thermal energy at room temperature. The reason for the opening of
a band gap is the broken inversion symmetry in the presence of WS$_2$. SOC lifts the
spin degeneracy at the K and K$'$ points and results in spin splittings of 8 meV
and 32 meV in the valence band of silicene and germanene, respectively, and 3 meV
and 19 meV in the conduction band, see Fig.\ \ref{fig:fig2}(b/d). The effect of the
SOC is enhanced in the presence of WS$_2$ as a consequence of tiny hybridization between
the Si/Ge $p$ and W $d$ orbitals. WS$_2$ indeed is characterized by very strong SOC, as
reflected by spin splittings of 431 meV (valence band) and 29 meV (conduction band) in
a freestanding monolayer \cite{epl}. The enhancement mechanism by proximity SOC is
similar to the case of graphene on transition metal dichalcogenides \cite{origin2}, but
due to the buckling of silicene/germanene the magnitude of spin splitting here is
different in the valence and conduction bands. As a consequence of the lifted spin
degeneracy, we obtain reduced band gaps of 23 meV and 14 meV for silicene and germanene
on WS$_2$, respectively.

\begin{figure}[t]
\centering
\includegraphics[width=0.5\textwidth,clip=true]{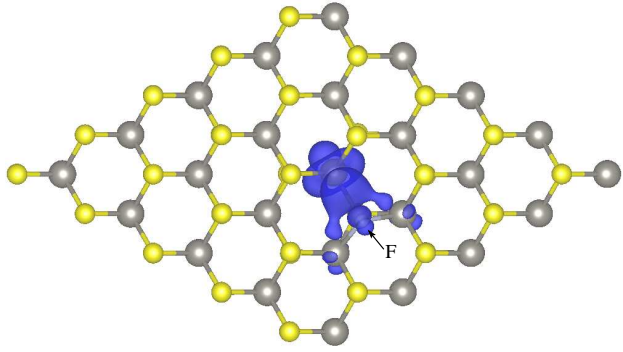}
\caption{Spin density in F-doped WS$_2$.}
\label{fig:fig3}
\end{figure}

In the following we will argue that silicene/germanene on F-doped WS$_2$ develops
energetically inequivalent band edges at the K and K$'$ points, i.e., spin-valley
polarization. The advantage of doping the substrate instead of the electronically active
material (the material giving rise to the states close to the Fermi energy) is that
impurity scattering is avoided. Specifically, we replace one S atom in the $5\times5\times1$
supercell of WS$_2$ with an F atom, corresponding to a doping concentration of 2\%,
in order to simulate the dilute doping limit. An earlier study has shown that 
substitutional F doping at the S site is possible in MoS$_2$ \cite{doping_prb}. We
calculate the binding energies $E({\rm WS_2, vac})+E({\rm X})-E({\rm WS_2, X})$
and obtain values of 4.3 eV for X = F and 6.1 eV for X = S, where $E({\rm WS_2, vac})$
is the energy of a relaxed WS$_2$ monolayer with one S vacancy, $E({\rm X})$ is the
energy of an X atom, and $E({\rm WS_2, X})$ is the energy of a relaxed WS$_2$ monolayer
with one S atom replaced by an X atom. Since the binding energies of F and S are
similar, it is likely that substitutional F doping at the S site is also possible in
WS$_2$. Experimental support for this conclusion comes from the
realization of P doping in MoS$_2$ and WSe$_2$ \cite{b1,b2} and Cl doping in MoS$_2$
and WS$_2$ \cite{b3}. By comparing spin degenerate and polarized
calculations for the doped supercell, we obtain an energy gain of 134 meV in the spin
polarized case and a total magnetic moment of 1 $\mu_B$. Next to the F atom the charge
transfer from the three neighbouring W atoms is reduced from two electrons to one
electron (F$^{-1}$ state instead of S$^{-2}$ state). The remaining electron is located
on one of the three W atoms and gives rise to the spatial distribution of spin density
shown in Fig.\ \ref{fig:fig3}. We obtain a small F magnetic moment of 0.05 $\mu_B$,
which is also reflected by Fig.\ \ref{fig:fig3}.

\begin{figure}[t]
\includegraphics[width=1.0\textwidth,clip=true]{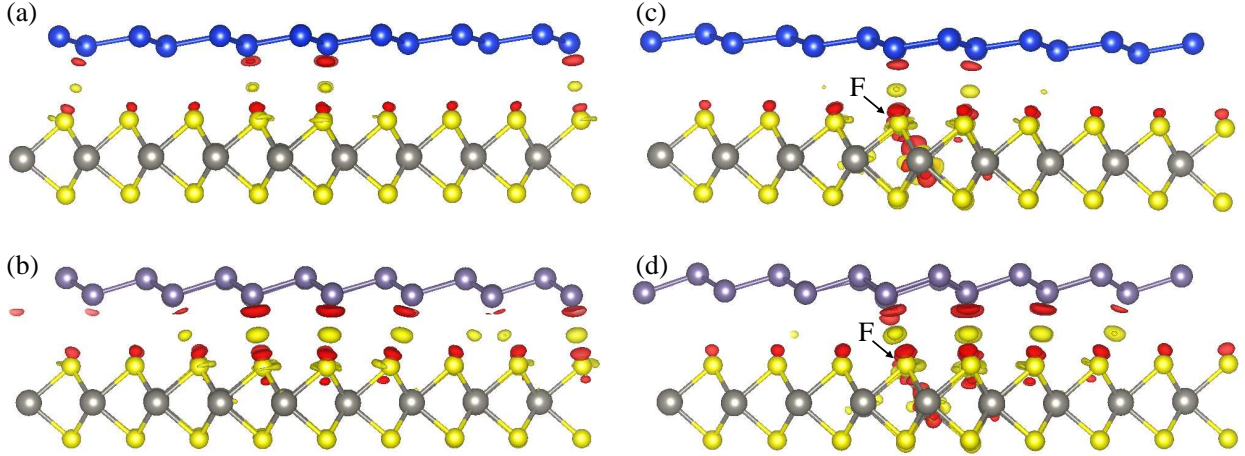}
\caption{Charge density difference induced by interaction of (a/b)
silicene/germanene with monolayer WS$_2$. Corresponding results for interaction with
F-doped WS$_2$ are given in (c/d). Yellow and red colors represent charge accumulation
and depletion, respectively. The plot takes into account the Si/Ge $p$,
W $d$, and S $p$ states, showing the $9\times10^{-4}$ electrons/Bohr$^3$ isosurface.}
\label{fig:fig4}
\end{figure}

\begin{figure}[t]
\centering
\includegraphics[width=0.5\textwidth,clip=true]{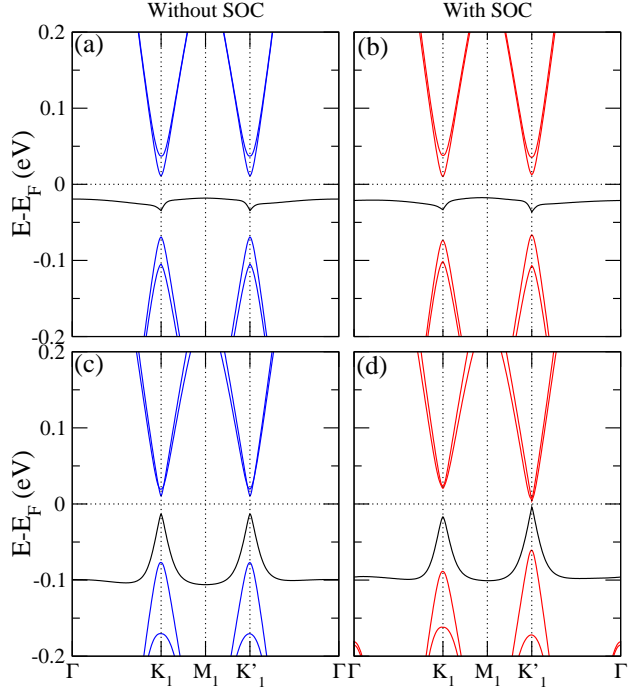}
\caption{Band structures of (a/b) silicene and (c/d) germanene on F-doped
monolayer WS$_2$. The SOC is neglected (left) or taken into account (right). The
F impurity band is shown in black color.}
\label{fig:fig5}
\end{figure}

We obtain for silicene and germanene on top of F-doped WS$_2$ smaller distances between
the component materials, 3.06 \AA\ and 2.79 \AA, respectively, as compared to the case
of pristine WS$_2$. This fact can be attributed to the additional magnetic coupling.
A comparison of the interaction in the cases of pristine and F-doped WS$_2$ is given
in Fig.\ \ref{fig:fig4} in terms of charge density difference plots. While the charge
redistribution in the van der Waals gap is only slightly modified after F doping,
compare the top to the bottom row of Fig.\ \ref{fig:fig4}, the dopant atom is strongly
affected, supporting the idea of magnetic coupling. We obtain still a total magnetic
moment of 1 $\mu_B$, carried largely by one W atom. This atom realizes a larger
distance (2.70 \AA) to the F atom than the other two neighbouring W atoms (2.30 \AA),
very similar to the situation without silicene/germanene. In addition, the F magnetic
moment is reduced to $0.05$ $\mu_B$. The band structure of silicene/germanene shows
significant alterations in contact with F-doped WS$_2$, see Fig.\ \ref{fig:fig5}.
Without SOC, the spin splitting in the valence band is enhanced to 36/91 meV and that
in the conduction band to 25/9 meV. This fact is expected to simply reduce the band
gaps, however, we observe in both cases the creation of an in-gap band just below the
Fermi energy. Analysis of the orbital character of this band shows that it arises almost
purely from the F impurity.

\begin{figure}[t]
\centering
\includegraphics[width=0.5\textwidth,clip=true]{fig6.eps}\\[0.3cm]
\includegraphics[width=0.3\textwidth,clip=true]{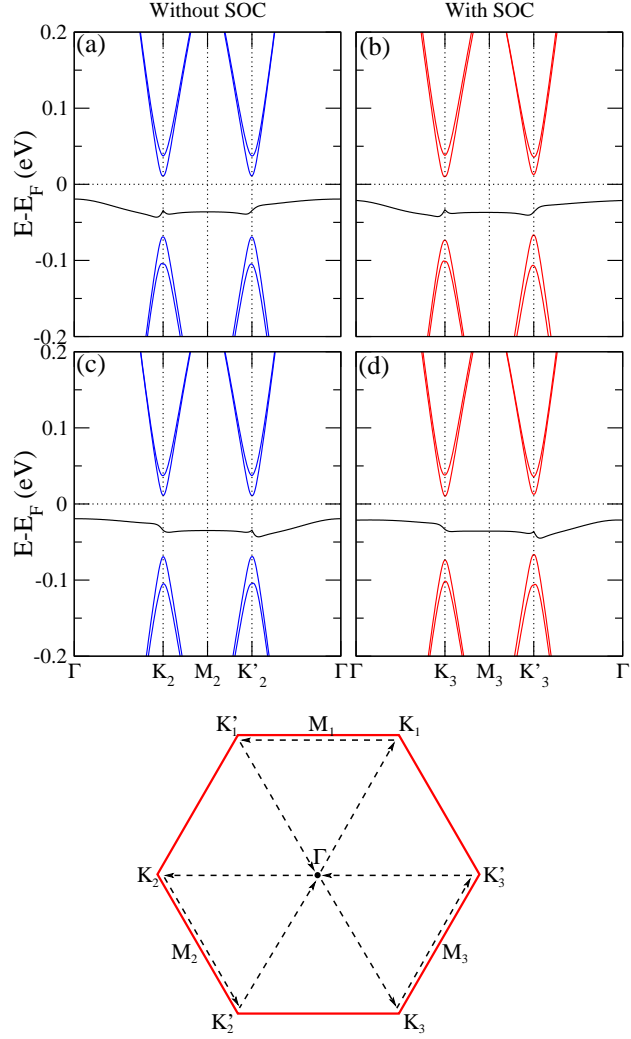}
\caption{Band structures of silicene on F-doped monolayer WS$_2$ for the 2 K and 2 K$'$
points not covered by Fig.\ \ref{fig:fig5}(a/b). The SOC is neglected (left) or taken
into account (right). The F impurity band is shown in black color. \textcolor{red}{
The different Brillouin zone paths are highlighted by dashed arrows (bottom).}}
\label{fig:fig6}
\end{figure}

Broken time-reversal symmetry due to the spin polarization in F-doped WS$_2$ induces
spin-valley polarization in silicene/germanene when the SOC is taken into account, see
Fig.\ \ref{fig:fig5}(b/d). Specifically, the spin splitting in the valence/conduction
band (not counting the F impurity band) is smaller/larger at the K than at the
K$'$ point. For silicene we obtain values of 28 meV and 40 meV for the valence band and
27 and 24 meV for the conduction band, respectively. More importantly, valley
polarization of 7 meV is created in the silicene valence band and one of 2 meV in the
conduction band. For germanene these effects are enhanced, with spin splittings of
74 meV and 111 meV (valence band) as well as 4 meV and 5 meV (conduction band) at the
K and K$'$ points, respectively. The valley polarization here amounts to 28 meV and
16 meV for the germanene valence and conduction bands, respectively, which opens a
route to spin-valley polarization even at room temperature.
In order to demonstrate that the K and K$'$ valleys are virtually not affected by the 
symmetry breaking in F-doped WS$_2$, as they belong almost purely to silicene/germanene,
we show in Fig.\ \ref{fig:fig6} for the silicene system the band structures of the
2 K and 2 K$'$ points not covered by Fig.\ \ref{fig:fig5}(a/b). We observe no lifting
of the valley degeneracy in the case without SOC and exactly the same valley structure
as before in the case with SOC. We also have checked that the spin hybridization at
the valleys is negligible. Moreover, while the F impurity band is located next to the
Fermi energy, the F states are spatially separated from the Si/Ge states and therefore
do not limit exploitation of the spin-valley polarization.

In conclusion, we have demonstrated that silicene and germanene in proximity to
monolayer WS$_2$ develop sizeable band gaps (due to inversion symmetry breaking) and
spin splittings (due to proximity SOC). F doping of WS$_2$ results in spin polarization
with a total magnetic moment of 1 $\mu_B$, since the charge transfer from the three
neighbouring W atoms is reduced. The remaining spin polarized electron is largely located
on one W atom. Our main finding is that F-doped WS$_2$ makes it possible to achieve
substantial spin-valley polarization in silicene and germanene. In contrast
to silicene, germanene even can enable valleytronics devices operating at ambient
conditions.

\begin{acknowledgments}
The research reported in this publication was supported by funding from King Abdullah
University of Science and Technology (KAUST).
\end{acknowledgments}

\end{document}